\documentclass{aa}
\usepackage{graphicx}
\usepackage{latexsym}
\usepackage{amsmath}
\usepackage{amssymb}
\usepackage[figuresleft]{rotating}

\newcommand{\MC}{\multicolumn}
\newcommand{\kms}{km~s$^\mathrm{-1}$}
\newcommand{\sunn}{$_{\odot}$}
\newcounter{qub}
\begin{document}

\title{Andromeda IV: a new Local Volume very metal-poor galaxy}

\author{%
S.A.~Pustilnik\inst{1} \and
A.L.~Tepliakova\inst{1} \and
A.Y.~Kniazev\inst{2,1}    \and
A.N.~Burenkov\inst{1}
}

\offprints{S.~Pustilnik  \email{sap@sao.ru}}

\institute{
Special Astrophysical Observatory RAS, Nizhnij Arkhyz,
Karachai-Circassia,  369167 Russia
\and South African Astronomical Observatory, Cape Town, South Africa
}

\date{Received 27 December 2007; Accepted \hskip 2cm  2008}

\abstract{
And~IV is a low-surface brightness (LSB) dwarf galaxy at the distance of
6.1 Mpc, projecting close to M~31. In this paper the results of spectroscopy
of And~IV the two brightest HII regions with the SAO 6-m telescope (BTA)
are presented. In both of them the faint line [OIII]$\lambda$4363~\AA\ was
detected that allowed us to determine their O/H by the classical T$_{\rm e}$
method. Their values of 12+$\log$(O/H) are equal to 7.49$\pm$0.06 and
7.55$\pm$0.23, respectively.
The comparison of these direct O/H determinations with the two most reliable
semi-empirical and empirical methods shows their good consistency.
For And~IV absolute blue magnitude of M$_{\rm B}$=--12.6, our value of O/H
corresponds well to the `standard' relation between O/H and L$_{\rm B}$ for
dwarf irregular galaxies (DIGs).
And~IV appears to be a new representative of the extremely metal-deficient
gas-rich galaxies in the Local Volume. The very large range of M(HI) for LSB
galaxies with close metallicities and luminosities indicates that the simple
models of LSBG chemical evolution are too limited to predict such striking
diversity.
}

\authorrunning{S.A.~Pustilnik et al.}

\titlerunning{And IV: a new Local Volume XMD LSB dwarf}

\maketitle

\section{Introduction}
\label{intro}

The very metal-poor galaxies were discovered more than 35 years ago. The
first and the most extreme of them was I~Zw~18 (Searle \& Sargent
\cite{SS72}). This galaxy, as well as the great majority of about five
dozen known to date late-type galaxies with the ISM value of
12+$\log$(O/H)$\leq$7.65 (or Z(ISM)$\leq$Z\sunn/10)\footnote{The solar value
of 12+$\log$(O/H), according to Asplund et al. (\cite{Solar04}), is accepted
to be equal to 8.66.}
(called eXtremely Metal-Deficient, or XMD galaxies. See, e.g., review
by Kunth \& \"Ostlin \cite{Kunth2000}) are
blue compact galaxies (BCGs). BCGs are low-mass galaxies with intense current
starburst, distinguished by strong emission-line spectra typical of HII
regions. They represent a small fraction of late-type low-mass galaxies in
the local Universe (a few percent). XMD BCGs, in turn, comprise only
$\sim$2\% of known BCGs (e.g., Pustilnik et al. \cite{Pustilnik_Kiel};
Kniazev et al. \cite{Kniazev03}).
The main interest to this small group is related to the similarity of their
properties to those predicted for high-redshift young galaxies.

Some of these XMD BCGs can themselves be rather young galaxies, with ages
of their oldest stars less than 1--2 Gyr. These include Tol~65 (Papaderos
et al. \cite{Papa99}), SBS 0335--052 E,W (Papaderos et al. \cite{Papa98},
Pustilnik et al. \cite{BTA}) and DDO~68 (Pustilnik et al.
\cite{DDO68,DDO68_sdss}).
However, the majority of XMD BCGs are likely older objects related to more
quiescent progenitor low-mass galaxies such as dwarf irregular (DIGs) and
LSB dwarf galaxies.
The concrete variants of possible evolutionary path-ways of XMD BCGs
could be elucidated with their detailed multiwavelength studies.

There are also about 15 of DIGs and LSB dwarf galaxies among XMD objects.
The majority of them are situated in the Local and the nearby groups, or in
the general field of the Local Volume (D $<$ 10 Mpc).
They include Leo~A and GR~8 (van Zee et al. \cite{vZee06}), Sextans A and B
(Skillman et al. \cite{Skillman89}; Kniazev et al. \cite{Sextans}), UGC 4483
(Izotov \& Thuan \cite{IT02}), DDO~53 (Pustilnik et al.
\cite{Pustilnik_DDO53}), UGCA 292 (van Zee \cite{vZee00}), SDSS J1215+5223
(Kniazev et al. \cite{Kniazev03}), ESO489--56 (Ronnback \& Bergvall
\cite{RB95}).
All these XMD galaxies are at the distances from 0.7 to 5 Mpc. There are
three more XMD LSBDs at the distances from 6.5 to 25 Mpc. Predominantly
small distances of mentioned DIGs and LSBD galaxies (in comparison to those
of XMD BCGs) are due to a selection effect. It is currently possible to
obtain the good quality spectra of `modest' HII regions in such objects
only for sufficiently close distances.

Thanks to their proximity, such galaxies can be studied in more detail at
many wavelengths. Hence, the origin of their very low metallicity can be
understood better and then the same question can be addressed to XMD BCGs.
The number of known XMD LSB galaxies is only nine, of which only four
are at the distances closer than $\sim$6 Mpc. Therefore, the discovery of
one more such object is useful in order to address the group properties and
their possible diversities as the appearance of different evolutionary
path-ways or initial or environmental conditions. Here we report that
And~IV, a LSB dwarf in the Local Volume with unusually high ratio
M(HI)/L$_{\rm B} \sim$13 (Chengalur et al. \cite{Chengalur07}), is one more
XMD galaxy.

And~IV, discovered more than 30 years ago by van den Bergh (\cite{vdBergh72}),
is projected close to the position of M~31 (at 41.9\arcmin\ from M~31
centre) and during long time its nature was uncertain. It was identified as a
background dwarf galaxy not related to M~31 after the HST photometry of its
individual stars by FGW.
With its integrated blue magnitude $B_{\rm tot}$=16.6 and the distance of
6.1 Mpc (FGW, Karachentsev et al. \cite{CNG}), its absolute magnitude is
M$_{\rm B}$=$-$12.6.
With the central SB of $\mu^{0}_{\rm V} =23.3$ mag~arcsec$^{-2}$ it should
be assigned as a LSB dwarf. Due to the proximity of a bright star (10$^{m}$)
to the main body of the galaxy, its study is not easy. FGW obtained H$\alpha$
images of And~IV and identified six HII regions in the body.
For four of them they obtained spectra with 4.2-m WHT telescope,
covering the range of 3700 to 7200~\AA. For three the brightest HII regions
(No. 3, 4 and 6 on their nomenclature) they estimated oxygen abundances by
the empirical R$_{23}$ method. These abundances appeared close each to other,
with a mean of 12+$\log$(O/H)=7.90. We reanalysed their spectra, adding to
their list of emission lines the faint line [OIII]$\lambda$4363~\AA, visible
in their plots for spectra of HII regions No. 3 and 4. The resulting O/H,
derived via the classical T$_{\rm e}$ method, appeared about a factor of two
lower (albeit with rather large uncertainty) than that derived by FGW.
To elucidate the real metallicity of And~IV and to improve the estimate of
its O/H, we conducted the spectroscopy of these HII regions with the SAO 6-m
telescope (BTA).

The paper is organised as follows.
In Sect.~\ref{Obs} we describe observations, obtained data and their
reduction. Sect.~\ref{results}
presents our results. Sect.~\ref{discussion} suggests the discussion of
And~IV as a new XMD galaxy and conclusions.

\section{Observational data and reduction}
\label{Obs}

\subsection{Observations}

The long-slit spectral observations were conducted with the SCORPIO
multimode instrument (Afanasiev \& Moiseev \cite{SCORPIO})
installed in the BTA prime focus, during the nights of January 12, 2007
and January 12, 2008. The grism VPHG550G was used with the 2K$\times$2K CCD
detector EEV~42-40. This allowed to registrate the range of
$\sim$3500--7500~\AA\ with the scale of $\sim$2.1~\AA~pixel$^{-1}$ and
the instrumental profile width of FWHM$\sim$12~\AA.
The scale along the slit, after binning, was 0\farcs357~pixel$^{-1}$, with
the total extent of $\sim$6\arcmin. On January 12, 2007 the slit (with width
of 1.0\arcsec) was positioned along the direction connecting the two
HII regions of interest (No.~3 and 4), that corresponded to the PA of
--26\degr. Parallactic angles during the exposure time varied between
60\degr\ and 67\degr\ and the zenith distances -- between 38\degr\ and
52\degr. Six 15-min exposures were obtained under the seeings of
FWHM$\sim$2.5--2.9\arcsec.
For observations on January 12, 2008 the long slit crossed only the brightest
HII-region No.~3 at PA=\mbox{PA=--132.5\degr}. Parallactic angles during the
exposure time varied between between 63\degr\ and 58.6\degr, while the
zenith distances -- between 48\degr\ and 56\degr. In the latter case the
slit orientation was rather close to the direction along atmospheric
refraction, so its effect was insignificant. Four 15-min exposures were
obtained under the seeing of FWHM$\sim$1.3\arcsec. Both nights were
photometric.
The object spectra were complemented before and after by the reference
spectra of He--Ne--Ar lamp for the wavelength calibration.
Bias and flat-field images were also acquired to perform the standard
reduction of 2D spectra. Spectrophotometric standard stars Feige~110,
G191B2B and Feige~34 (Bohlin \cite{Bohlin96}) were observed during the
night for the flux calibration.

\subsection{Reduction}

\begin{table*}[hbtp]
\centering{
\caption{Line intensities and the derived parameters in And~IV HII
regions No.~3 and 4}
\label{t:Intens}
\begin{tabular}{|l|c|c|c|c|c|c|} \hline  \hline
\rule{0pt}{10pt}
\rule{0pt}{10pt}
	    &   \MC {2}{c|}{No.~3, 12.01.07} & \MC {2}{c|}{No.~3, 12.01.08}  &\MC {2}{c|}{No.~4, 12.01.07}  \\ \hline
$\lambda_{0}$(\AA) Ion                    &
$F$($\lambda$)/$F$(H$\beta$)&$I$($\lambda$)/$I$(H$\beta$) & $F$($\lambda$)/$F$(H$\beta$)&$I$($\lambda$)/$I$(H$\beta$) & $F$($\lambda$)/$F$(H$\beta$)&$I$($\lambda$)/$I$(H$\beta$) \\ \hline

3727\ [O\ {\sc ii}]\            & 1.744$\pm$0.156 & 1.723$\pm$0.165  & 1.419$\pm$0.079 & 1.669$\pm$0.097     & 2.166$\pm$0.274 & 2.427$\pm$0.321  \\
3869\ [Ne\ {\sc iii}]\          & 0.173$\pm$0.048 & 0.170$\pm$0.048  & 0.139$\pm$0.046 & 0.160$\pm$0.054     &      ...        &      ...         \\
3889\ He\ {\sc i}\ + \ H8\      & 0.144$\pm$0.027 & 0.167$\pm$0.042  & 0.192$\pm$0.049 & 0.242$\pm$0.064     &      ...        &      ...         \\
3967\ [Ne\ {\sc iii}]\ +\ H7\   & 0.089$\pm$0.019 & 0.114$\pm$0.039  & 0.116$\pm$0.008 & 0.157$\pm$0.016     &      ...        &      ...         \\
4101\ H$\delta$\                & 0.198$\pm$0.028 & 0.217$\pm$0.039  & 0.184$\pm$0.011 & 0.227$\pm$0.018     & 0.287$\pm$0.085 & 0.311$\pm$0.116  \\
4340\ H$\gamma$\                & 0.489$\pm$0.040 & 0.501$\pm$0.046  & 0.448$\pm$0.015 & 0.494$\pm$0.019     & 0.459$\pm$0.074 & 0.484$\pm$0.132  \\
4363\ [O\ {\sc iii}]\           & 0.062$\pm$0.015 & 0.062$\pm$0.015  & 0.043$\pm$0.008 & 0.046$\pm$0.009     & 0.094$\pm$0.050 & 0.098$\pm$0.053  \\
4471\ He\ {\sc i}\              &      ...        &     ...          & 0.019$\pm$0.006 & 0.020$\pm$0.007     &      ...        &      ...         \\
4861\ H$\beta$\                 & 1.000$\pm$0.077 & 1.000$\pm$0.079  & 1.000$\pm$0.027 & 1.000$\pm$0.028     & 1.000$\pm$0.098 & 1.000$\pm$0.146  \\
4959\ [O\ {\sc iii}]\           & 0.664$\pm$0.055 & 0.656$\pm$0.055  & 0.699$\pm$0.055 & 0.680$\pm$0.020     & 0.936$\pm$0.097 & 0.925$\pm$0.096  \\
5007\ [O\ {\sc iii}]\           & 2.132$\pm$0.053 & 2.059$\pm$0.052  & 2.132$\pm$0.053 & 2.059$\pm$0.052     & 3.050$\pm$0.268 & 3.000$\pm$0.266  \\
5876\ He\ {\sc i}\              & 0.101$\pm$0.023 & 0.100$\pm$0.023  & 0.074$\pm$0.007 & 0.064$\pm$0.006     &      ...        &      ...         \\
6548\ [N\ {\sc ii}]\            & 0.015$\pm$0.013 & 0.015$\pm$0.014  & 0.008$\pm$0.006 & 0.006$\pm$0.005     & 0.003$\pm$0.017 & 0.002$\pm$0.015  \\
6563\ H$\alpha$\                & 2.430$\pm$0.183 & 2.408$\pm$0.200  & 3.426$\pm$0.082 & 2.793$\pm$0.074     & 3.122$\pm$0.275 & 2.744$\pm$0.267  \\
6584\ [N\ {\sc ii}]\            & 0.050$\pm$0.026 & 0.049$\pm$0.026  & 0.027$\pm$0.027 & 0.022$\pm$0.023     & 0.019$\pm$0.036 & 0.017$\pm$0.032  \\
6717\ [S\ {\sc ii}]\            & 0.104$\pm$0.017 & 0.103$\pm$0.017  & 0.101$\pm$0.009 & 0.081$\pm$0.008     & 0.166$\pm$0.040 & 0.144$\pm$0.036  \\
6731\ [S\ {\sc ii}]\            & 0.036$\pm$0.014 & 0.035$\pm$0.014  & 0.087$\pm$0.010 & 0.070$\pm$0.008     & 0.022$\pm$0.033 & 0.019$\pm$0.029  \\  \hline
C(H$\beta$)\ dex          & \MC {2}{c|}{0.00$\pm$0.10}    & \MC {2}{c|}{0.25$\pm$0.03}  & \MC {2}{c|}{0.16$\pm$0.11}  \\
EW(abs)\ \AA\             & \MC {2}{c|}{1.65$\pm$1.78}    & \MC {2}{c|}{2.30$\pm$1.01}  & \MC {2}{c|}{0.35$\pm$11.6}  \\
$F$(H$\beta$)$^a$\        & \MC {2}{c|}{ 14.51$\pm$1.43}  & \MC {2}{c|}{ 16.88$\pm$0.29}& \MC {2}{c|}{3.62$\pm$0.25}   \\
EW(H$\beta$)\ \AA\        & \MC {2}{c|}{ 136$\pm$7 }      & \MC {2}{c|}{ 150$\pm$7 }    & \MC {2}{c|}{ 108$\pm$8}   \\
\hline  \hline
\MC{7}{l}{$^a$ in units of 10$^{-16}$ ergs~s$^{-1}$cm$^{-2}$.}\\
\end{tabular}
}
\end{table*}

The standard pipeline with the use of IRAF\footnote{IRAF: the Image Reduction
and Analysis Facility is
distributed by the National Optical Astronomy Observatory, which is
operated by the Association of Universities for Research in Astronomy,
In. (AURA) under cooperative agreement with the National Science
Foundation (NSF).}
and MIDAS\footnote{MIDAS is an acronym for the European Southern
Observatory package -- Munich Image Data Analysis System. }
was applied for the reduction of the long-slit spectra, which included the
next steps.

Cosmic ray hits were removed from the 2D spectral frames in MIDAS.
Using IRAF packages from CCDRED, we subtracted bias and performed
flat-field correction.
After that the 2D spectra were wavelength calibrated 
and the night sky background was subtracted.
Then, using the data on the spectrophotometry of standard stars,
the 2D spectra were transformed to absolute fluxes.
After that, using the continuum of the brightest HII region, all spectra
were straighten. Finally, for each individual 2D spectrum a similar
subregion was separated along the slit, centred at the maximum of continuum
in the spectrum of HII region No.3, and all 6 individual 2D spectra were
summed up (without weighting) in order to obtain the full time exposure
2D spectrum.
One-dimensional spectra of the central parts of HII regions were then
extracted by
summing up, without weighting, 18 and 6 rows along the slit ($\sim$6.3 and
2.1\arcsec) for HII region 3 for observations in Jan. 2007 and Jan. 2008,
respectively, and 9 rows ($\sim$3.1\arcsec) - for HII region 4. These sizes
corresponded to the subregions, where the principal line [OIII]$\lambda$4363
for determination of T$_{\rm e}$, was `visible' above the noise.
The significant difference in the sizes of exctracted subregions for HII
region No.~3 is related with two factors. First, as it is seen on the HST
image from FGW, this region is well elongated along nearly the North--South
direction, with visible extent of $\sim$4.5\arcsec$\times$$\sim$2.5\arcsec.
For the observation in Jan.~2007 the long slit was oriented approximately
along the major axis, while for Jan.~2008 -- approximately along the minor
axies. Second, the seeings for these runs differed significantly:
$\sim$2.7\arcsec\ in Jan.~2007 versus 1.3\arcsec\ in Jan.~2008.

All emission lines were measured applying the MIDAS programs
described in detail in Kniazev et al. (\cite{SHOC}).
Briefly, they draw continuum, perform robust noise estimation and fit
separate lines by a single Gaussian superimposed on the continuum-subtracted
spectrum. Emission lines blended in pairs or triplets were fitted
simultaneously as blend of two or three Gaussians features. The quoted errors
of singular line intensities include the following components. The first is
related to the Poisson statistics of line photon flux. The second component
is the error resulting from the creation of the underlying continuum, which
gives the main contribution for the errors of faint lines. For fluxes of
lines in blends an additional error appears related to the goodness of fit.
Last, the term related to the uncertainty of the spectral sensitivity curve
gives an additional error to the relative line intensities. For observations
in Jan. 2007 the half-amplitude of variations of sensitivity curve along
the whole working range ($\lambda >$ 3650~\AA) was less than
1\%. For observation in Jan. 2008 it was less than 0.5~\%.
All these components are summed up squared. The total errors have been
propagated to calculate the errors of all derived parameters.

\section{Results}
\label{results}

\subsection{Line intensities and element abundances}
\label{abun}

\begin{figure*}
   \centering
 \includegraphics[angle=-90,width=15.0cm, clip=]{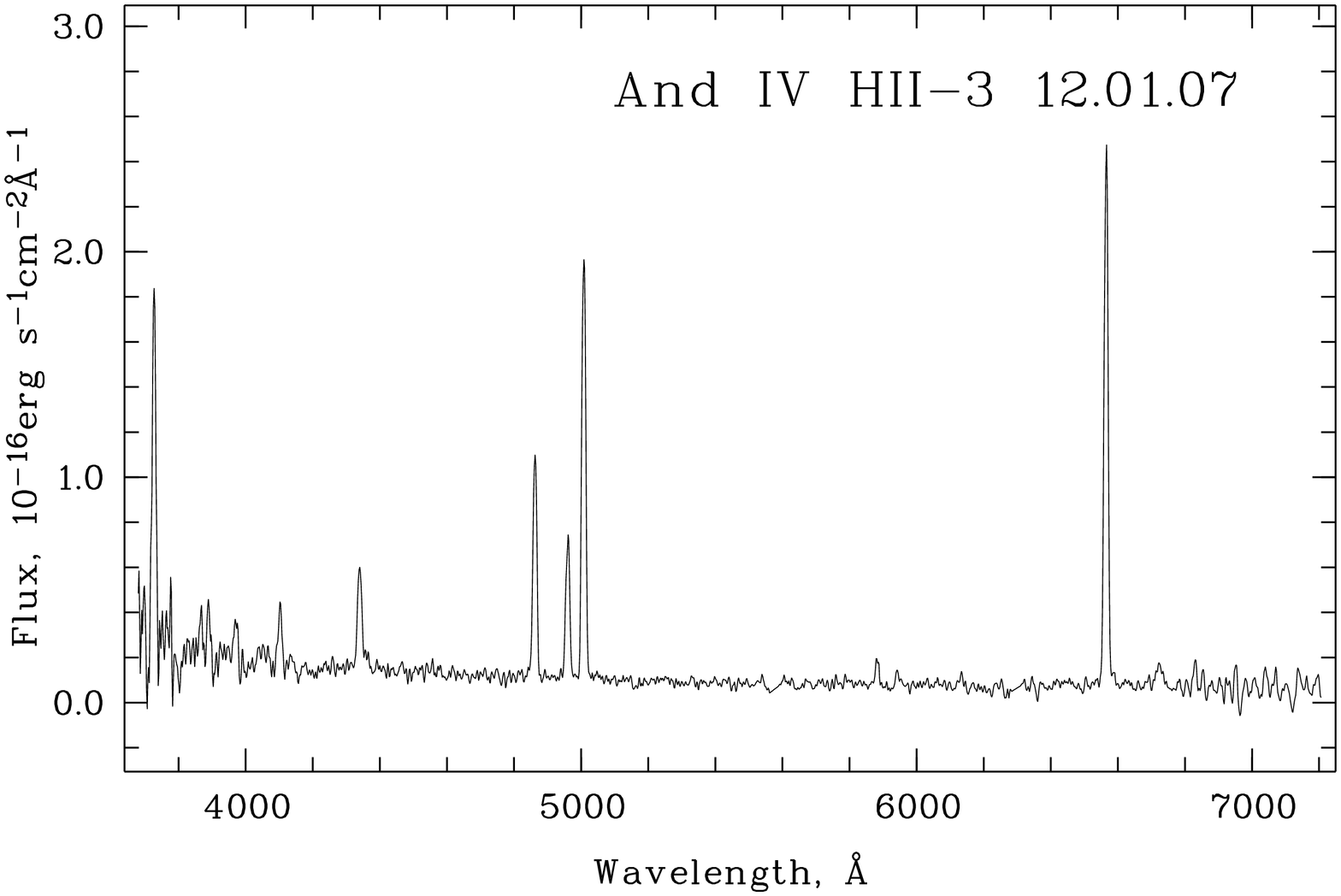}
 \includegraphics[angle=-90,width=15.0cm, clip=]{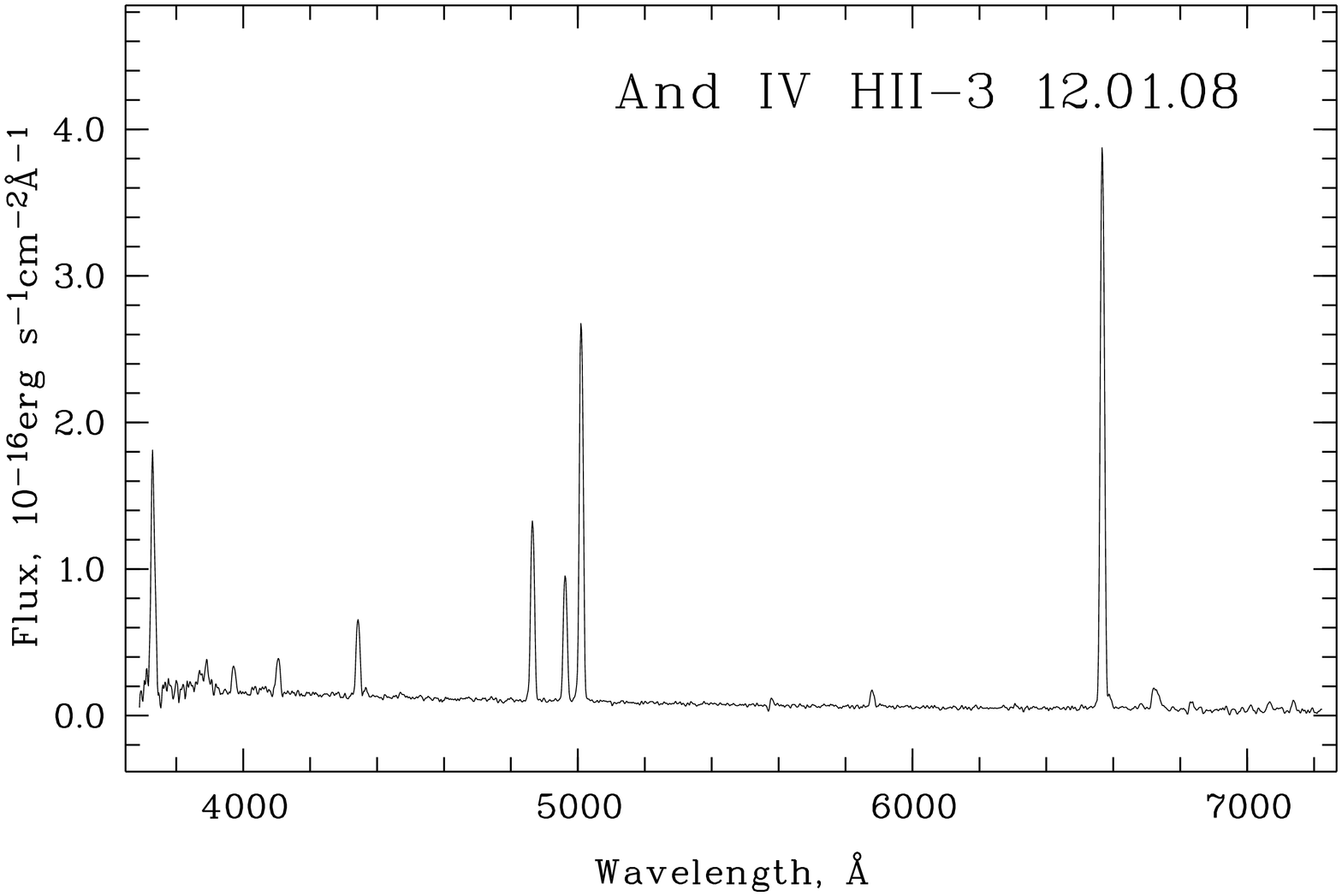}
   \caption{
   The BTA spectra of HII region No.3, obtained on 12.01.07 (top panel)
   and 12.01.08 (bottom panel). Numbers are according to FGW nomenclature.
       }
	 \label{fig:spectra_HIIR3}
 \end{figure*}

\begin{figure*}
   \centering
 \includegraphics[angle=-90,width=15.0cm, clip=]{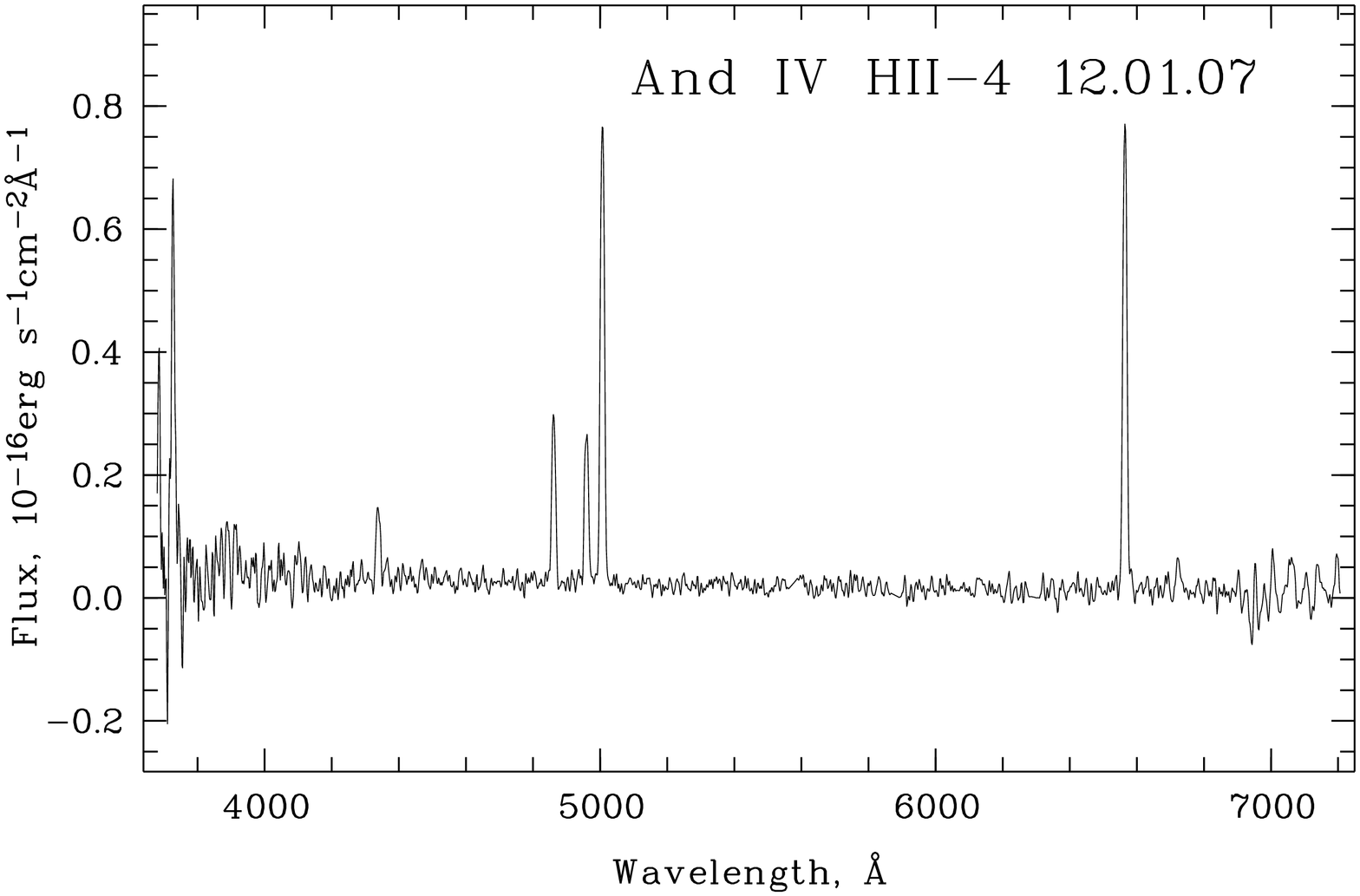}
   \caption{
   The BTA spectrum of HII region No.4, obtained on 12.01.07.
       }
	 \label{fig:spectra_HIIR4}
 \end{figure*}

Relative intensities of all emission lines used for the abundance
determination in the HII regions of And~IV,
as well as the derived C(H$\beta$), the EWs of Balmer absorption lines, the
measured flux in H$\beta$ emission line and the measured heliocentric
radial velocity are given in Table \ref{t:Intens}.
Their spectra are shown in
Figures~\ref{fig:spectra_HIIR3}, \ref{fig:spectra_HIIR4}.
Extinction is low: C(H$\beta$) $\lesssim$0.1,
consistent with the extinction value in our Galaxy and with data for most
very metal-poor galaxies.

To derive the element  abundances of species O, Ne, N in the HII regions of
And~IV, we use  the standard method from Aller (\cite{Aller84}), and follow
the procedure described in detail by Pagel et al. (\cite{Pagel92}) and
Izotov et al. (\cite{ITL97}).
Element abundances and physical parameters are determined in the frame of
the classical  two-zone model of HII region (Stasi\'nska
\cite{Stas90}), as described in detail in our papers (Pustilnik
et al. \cite{HS0837},  Kniazev et al. \cite{SHOC}; \cite{Sextans}).
The derived  electron temperatures T$_{\rm e}$ and density N$_{\rm e}$,
as well as the abundances of oxygen, nitrogen, and neon
are given in Table~\ref{t:Chem}. The resulting oxygen abundances, derived
for region No.~3 in January 2007 and 2008, correspond to the value of
parameter 12+$\log$(O/H) equal to 7.40$\pm$0.11 and 7.56$\pm$0.08,
respectively, and for region No.~4 -- to 7.55$\pm$0.23.
The weighted mean on two independent measurements of O/H in region No.~3
(derived on O/H and its error in Table~\ref{t:Chem}) corresponds to
12+$\log$(O/H)=7.49$\pm$0.06. The value of O/H for region No.~4, despite
it has a much larger uncertainty, is well consistent with the weighted mean
for region No.~3.

The latter value of O/H allows to classify this dwarf LSB galaxy  as one of
the most metal-poor galaxies known in the Local Volume. The weighted mean
abundance ratios N/O ($\log$(N/O)=--1.78$\pm$0.15) and Ne/O
($\log$(Ne/O)=--0.77$\pm$0.12) for HII region No.3, within their cited errors
are close to the typical of XMD BCGs (Izotov \& Thuan \cite{IT99}). This
seemingly indicates common properties of heavy element enrichment in both
types of
galaxies, and hence sufficiently similar IMFs (Initial Mass Functions) in
the range of large and intermediate stellar masses.

\begin{table*}[hbtp]
\centering{
\caption{The O, N and Ne abundances in HII regions No.~3 and 4 of And~IV}
\label{t:Chem}
\begin{tabular}{|l|c|c|c|} \hline  \hline
					   & No.~3, 12.01.07  & No.~3, 12.01.08     & No.~4, 12.01.07    \\  \hline
$T_{\rm e}$(OIII)(10$^{3}$~K)\             & 19.38$\pm$2.79   & 16.02$\pm$1.54      & 19.89$\pm$6.51     \\
$T_{\rm e}$(OII)(10$^{3}$~K)\              & 15.50$\pm$2.08   & 14.22$\pm$1.30      & 15.65$\pm$4.72     \\
$N_{\rm e}$(SII)(cm$^{-3}$)\               &  10$\pm$10~~     & 316$\pm$288~        &  10$\pm$10~~       \\
&    &   & \\
O$^{+}$/H$^{+}$($\times$10$^{-5}$)\        & 1.356$\pm$0.476  & 1.757$\pm$0.453     & 1.856$\pm$1.429    \\
O$^{++}$/H$^{+}$($\times$10$^{-5}$)\       & 1.183$\pm$0.396  & 1.905$\pm$0.459     & 1.709$\pm$1.272    \\
O/H($\times$10$^{-5}$)\                    & 2.538$\pm$0.620  & 3.662$\pm$0.645     & 3.654$\pm$1.906    \\
12+$\log$(O/H)\                            & ~7.40$\pm$0.11~  & ~7.56$\pm$0.08~     & ~7.55$\pm$0.23~    \\
&   &  &  \\
 N$^{+}$/H$^{+}$($\times$10$^{-6}$)\       & 0.343$\pm$0.159~~& 0.178$\pm$0.146~~   &    ...               \\
 ICF(N)\                                   &  1.872           &  2.084              &    ...                \\
 $\log$(N/O)\                              & --1.60$\pm$0.23~~& --1.99$\pm$0.36~~   &    ...               \\
&  &  &  \\
Ne$^{++}$/H$^{+}$($\times$10$^{-5}$)\      & 0.213$\pm$0.089  & 0.323$\pm$0.134     &     ...            \\
ICF(Ne)\                                   & 2.146            & 1.992               &     ...              \\
$\log$(Ne/O)\                              & --0.74$\pm$0.21~ & --0.77$\pm$0.20~    &     ...             \\
\hline   \hline
\end{tabular}
}
\end{table*}

\begin{table*}[hbtp]
\centering{
\caption{Line intensities in And~IV HII regions No.~3 and 4 according
to FGW and our derived C(H$\beta$) and EW(abs)}
\label{t:Intens_Fer}
\begin{tabular}{|l|c|c|c|c|} \hline  \hline
\rule{0pt}{10pt}
\rule{0pt}{10pt}
	    &   \MC {2}{c}{HII-region No.~3} & \MC {2}{c|}{HII-region No.~4}  \\ \hline
$\lambda_{0}$(\AA) Ion                    &
$F$($\lambda$)/$F$(H$\beta$)&$I$($\lambda$)/$I$(H$\beta$) &$F$($\lambda$)/$F$(H$\beta$)&$I$($\lambda$)/$I$(H$\beta$) \\ \hline

3727\ [O\ {\sc ii}]\   & 2.280$\pm$0.248 & 2.627$\pm$0.311 & 1.810$\pm$0.196 & 1.846$\pm$0.217  \\
4101\ H$\delta$\       & 0.290$\pm$0.034 & 0.312$\pm$0.047 & 0.300$\pm$0.034 & 0.306$\pm$0.044  \\
4340\ H$\gamma$\       & 0.450$\pm$0.046 & 0.478$\pm$0.054 & 0.490$\pm$0.054 & 0.495$\pm$0.058  \\
4363\ [O\ {\sc iii}]\  & 0.045$\pm$0.014 & 0.048$\pm$0.015 & 0.061$\pm$0.025 & 0.062$\pm$0.025  \\
4861\ H$\beta$\        & 1.000$\pm$0.111 & 1.000$\pm$0.112 & 1.000$\pm$0.111 & 1.000$\pm$0.112  \\
4959\ [O\ {\sc iii}]\  & 0.620$\pm$0.065 & 0.614$\pm$0.065 & 1.060$\pm$0.115 & 1.056$\pm$0.115  \\
5007\ [O\ {\sc iii}]\  & 2.060$\pm$0.227 & 2.028$\pm$0.224 & 3.050$\pm$0.337 & 3.038$\pm$0.337  \\
6548\ [N\ {\sc ii}]\   & 0.033$\pm$0.010 & 0.029$\pm$0.009 & 0.027$\pm$0.010 & 0.026$\pm$0.010  \\
6563\ H$\alpha$\       & 3.240$\pm$0.361 & 2.782$\pm$0.338 & 2.860$\pm$0.322 & 2.791$\pm$0.342  \\
6584\ [N\ {\sc ii}]\   & 0.100$\pm$0.014 & 0.086$\pm$0.012 & 0.080$\pm$0.012 & 0.078$\pm$0.013  \\
6717\ [S\ {\sc ii}]\   & 0.110$\pm$0.014 & 0.093$\pm$0.013 & 0.130$\pm$0.016 & 0.127$\pm$0.017  \\
6731\ [S\ {\sc ii}]\   & 0.080$\pm$0.012 & 0.068$\pm$0.011 & 0.090$\pm$0.013 & 0.088$\pm$0.014  \\  \hline
C(H$\beta$)\ dex       & \MC {2}{c|}{0.20$\pm$0.14} & \MC {2}{c|}{0.03$\pm$0.15}  \\
EW(abs)\ \AA\          & \MC {2}{c|}{0.00$\pm$2.73} & \MC {2}{c|}{1.20$\pm$11.66}  \\
$F$(H$\beta$)$^a$\     & \MC {2}{c|}{ 8.8        }  & \MC {2}{c|}{ 4.5        }   \\
EW(H$\beta$)\ \AA\     & \MC {2}{c|}{ 171       }   & \MC {2}{c|}{ 684       }   \\
Rad. vel.\ \kms\       & \MC {2}{c|}{ 244$\pm$15}   & \MC {2}{c|}{ 250$\pm$13}    \\
\hline  \hline
\MC{5}{l}{$^a$ in units of 10$^{-16}$ ergs\ s$^{-1}$cm$^{-2}$.}\\
\end{tabular}
 }
\end{table*}

\section{Discussion and conclusions}
\label{discussion}

\subsection{Comparison with other determinations}

The signal-to-noise ratios in the faint line [OIII]$\lambda$4363~\AA, that
is used for the determination of electron temperature in HII region, is rather
small in our spectra: \mbox{($\sim$2--$\sim$5)}. Therefore, independent
opportunities to get the estimates of O/H and compare them with the results
of the direct T$_{\rm e}$ method, of course are worse to employ.
There are several empiric methods, which use for O/H estimate the intensities
of strong lines of oxygen and other elements. Their review with the
estimates of their accuracy was given recently by Yin et al. (\cite{Yin07}).
Besides, Izotov and Thuan (\cite{IT07}) suggested a semi-empirical method, in
which T$_{\rm e}$ is estimated on sum of intensities of strong oxygen lines,
basing on relation found in Stasinska \& Izotov (\cite{Stas03}) from the
combined analysis of observational data and large grid of HII region models.
After the T$_{\rm e}$ is estimated by this empirical relation, all other
abundance calculations are performed exactly as in the classic method.
Izotov and Thuan (\cite{IT07}) compared for their sample of low-metallicity
galaxies (a dozen HII regions in eight galaxies) O/H derived by the direct
method, and those obtained by several empirical and their semi-empirical
methods. The least differences with the direct method derived O/H were
obtained for their semi-empirical method (rms=0.06~dex) and the empirical
formula for O/H suggested in paper by Yin et al. (\cite{Yin07})
(rms=0.07~dex):
$$ 12+\log(O/H) = 6.486 + 1.401\log(R_{23}), $$

where $$R_{23} = (I(3727) + I(4959) + I(5007))/I(H\beta). $$
These estimates we will use for comparison with direct determinations.
The estimates for parameter 12+$\log$(O/H), obtained for two observations of
region No.~3 (Jan.~2007 and Jan.~2008) and for region No.~4 by semi-empirical
method, give respectively the following numbers: 7.48, 7.49 É 7.76.
The estimates of the same parameters with the use of empirical formula from
Yin et al. (\cite{Yin07}) give the following numbers: 7.36, 7.39 É 7.61.

Thus, for HII-region No.~3 semi-empirical estimates of O/H for both
observations (7.48, 7.49) are very close to the weighted mean, determined
by the direct method on the same spectra (7.49). The estimates of O/H on
the empirical formula from Yin et al. (\cite{Yin07}) for region No.~3 are
well consistent each to other, but give the value of O/H which is
systematically lower by $\sim$0.1~dex. For region No.~4 both methods give
significantly higher values of O/H (7.76 É 7.61), which however are
consistent with the direct determination 7.55, accounting for its large
error of 0.23~dex.  Thus, the estimates with the help of the most reliable
semi- and empirical methods, confirms that region No.~3 has very low
metallicity. For region No.~4 the available data do not contradict that it
also has very low O/H. It is, however, not excluded, that its value is
somewhat higher than in region No.~3, by  $\sim$0.10--0.15~dex.

We can similarly use the spectral data for these regions from paper by FGW.
First, on the plots of the spectra of HII regions No.~3 and 4, presented
by FGW, there is marginally visible, but measurable line
[OIII]$\lambda$4363~\AA. We estimated its intensity on these plots based
on the relative heights of this line and
of the nearby H$\gamma$-line. To perform the abundance calculations through
the classical T$_{\rm e}$ method like to that for our BTA spectra, we
transformed their corrected for CH$\beta$ line intensities of HII region No.~3
(for which C(H$\beta$) in FGW is larger than zero) back to the measured line
fluxes. For a reader convenience, these, as well as the line fluxes for
region No.~4 are given in Table~\ref{t:Intens_Fer} with the accepted values
for the line [OIII]$\lambda$4363~\AA. The results of our calculations of
physical conditions and abundances are given in Table~\ref{t:Chem_Fer}.

The values of 12+$\log$(O/H) for HII regions No.~3 and 4 derived on these
data by the direct T$_{\rm e}$ method are equal to 7.63$\pm$0.14 and
7.72$\pm$0.16, respectively. The estimates by semi-empirical method give
very close O/H values, of 7.64 and 7.70, respectively. Using the empirical
formula from Yin et al. (\cite{Yin07}), we again get O/H systematically
smaller: 7.48 and 7.57, respectively. Thus, all our estimates of O/H on the
spectra of HII regions No.~3 and 4 from paper by FGW are consistent, within
the cited errors, with O/H values derived on our observations. There is
an indication that O/H in region No.~4 is somewhat higher than that in
region No.~3.

\begin{table}[hbtp]
\centering{
\caption{Our O and N abundances in HII regions No.~3 and 4 of
And~IV on the line intensities of FGW from Table~\ref{t:Intens_Fer}}
\label{t:Chem_Fer}
\begin{tabular}{|l|c|c|} \hline  \hline
				       & No.~3              & No.~4              \\  \hline
$T_{\rm e}$(OIII)(10$^{3}$~K)\         & 16.64$\pm$2.75     & 15.27$\pm$2.98     \\
$T_{\rm e}$(OII)(10$^{3}$~K)\          & 14.49$\pm$2.28     & 13.88$\pm$2.57     \\
$T_{\rm e}$(SIII)(10$^{3}$~K)\         & 15.51$\pm$2.28     & 14.37$\pm$2.47     \\
$N_{\rm e}$(SII)(cm$^{-3}$)\           &  45$\pm$226~       &  10$\pm$10~~       \\
&    & \\
O$^{+}$/H$^{+}$($\times$10$^{-5}$)\    & 2.535$\pm$1.117    & 2.028$\pm$1.083    \\
O$^{++}$/H$^{+}$($\times$10$^{-5}$)\   & 1.684$\pm$0.670    & 3.196$\pm$1.623    \\
O/H($\times$10$^{-5}$)\                & 4.219$\pm$1.318    & 5.224$\pm$1.952    \\
12+$\log$(O/H)\                        & ~7.63$\pm$0.14~    & ~7.72$\pm$0.16~    \\
&   & \\
 N$^{+}$/H$^{+}$($\times$10$^{-6}$)\    & 0.679$\pm$0.207~~  & 0.673$\pm$0.247~~    \\
 ICF(N)\                                &  1.664             &  2.576                \\
 $\log$(N/O)\                           & --1.57$\pm$0.19~~  & --1.48$\pm$0.23~~    \\
\hline   \hline
\end{tabular}
 }
\end{table}

The comparison of the relative line intensities in both HII regions,
measured in our work, with those from the paper by FGW, shows their general
consistency, with  except of I(H$\alpha$) in January 2007 and
I([OII]$\lambda$3727) in HII region No.3. I(H$\alpha$) is somewhat lower
than the expected theoretical value, but the difference can be treated to
be within cited errors. Our I([OII]$\lambda$3727) appears somewhat lower
than that in FGW.  The most likely this is related with subtracting
different parts of region No.~3 in our work and by FGW.
The effect of differential atmospheric refraction (see, e.g., Filippenko
\cite{Filippenko82}) was insignificant. In January 2008, when the seeing
was of FWHM=1.3\arcsec, the long slit orientation was close to the direction
of atmospheric refraction. For observations in January 2007 the slit was
oriented almost perpendicular to the latter. However, for the slit width
and average seeing for that case of FWHM=2.7\arcsec, the effect of
differential refraction is almost washed out.

Thus, summarising the comparison of our direct O/H determinations in the
two brightest HII regions of And~IV with their most reliable indirect
estimates, and also with the estimates, derived on the spectra of the same
HII regions from FGW, we conclude that all of them are well consistent with
our determinations. The empirical formula from Yin et al. (\cite{Yin07}
gives seemingly in these regions somewhat smaller O/H by about 0.1~dex.
Accounting for the large difference in the errors for O/H in regions No.~3
and 4 (7.49$\pm$0.06 and 7.55$\pm$0.23, respectively), we accept for further
the parameter 12+$\log$(O/H) in And~IV to be equal to 7.49$\pm$0.06.

\subsection{The L--Z relation and And~IV}

Since And~IV appeared a new XMD LSB dwarf galaxy, it is interesting to
examine how this fits in the well known metallicity-luminosity relation for
DIGs and related objects (Skillman et al. \cite{Skillman89}).
The updated version of L--Z relation in $B$-band, incorporating all
available data for DIGs in the LV was presented by Lee et al. (\cite{Lee03}).
Similar L--Z relation for an independent sample of DIGs was derived by van Zee
\& Haynes (2006).
One more similar relation for DIGs within the volume with D$<$5~Mpc was
obtained by van Zee et al. (\cite{vZee06}).
All three relations are very close each to other, but that of van Zee \&
Haynes (\cite{vZee_Haynes}) has the least rms scatter (0.15 dex), so
we use this relation: 12+$\log$(O/H)=5.65--0.149$\times$M$_{\rm B}$
to compare the derived O/H values for And~IV.
With the And~IV value of M$_{\rm B}$=--12.60 (Karachentsev et al.
\cite{CNG}), the value of O/H, expected from the latter L--Z relation is
7.53. This value is 0.37 dex (or 2.5$\times$r.m.s.) smaller than the O/H,
obtained by FGW. Our value of 12+$\log$(O/H)=7.49$\pm$0.06,
agrees with the L--Z relation very well.

Izotov et al. (\cite{Izotov05}) presented the updated formulae for
calculations of element abundances in HII regions, incorporating updates
in atomic parameters of relevant ions. This results in small systematic
differences with the results from the standard models, used in this work.
In particular, for O/H this difference changes approximately linearly
from +0.05~dex to +0.005 dex for T$_{\rm e}$ in the range of 10000 to 20000~K,
that is new formulae give a bit higher O/H values. Since many recent O/H
determinations (e.g., in Izotov \& Thuan \cite{IT07}) are given in the new
scale, for the correct comparison we give O/H for And~IV also in the new
scale.
For the respective range of T$_{\rm e}$ we need to add to O/H $\sim$0.01 dex.
Thus, in the new scale, And~IV, with 12+$\log$(O/H)=7.50$\pm$0.06, enters
to the five most metal-poor DIG/LSBDs in the Local Volume.

\subsection{And~IV and the diversity in properties of LSB dwarf galaxies}

One of the Introduction points was that related to the possibility to
study the diversity of dwarf galaxy properties in the Local Volume and their
differences in evolutionary path-ways. It is worth to notice some significant
differences between And~IV and other XMD LSBDs.  In particular, in
Table~\ref{tab:2_LSB}
we present for comparison the global parameters of And~IV and another nearby
XMD LSBD ESO~489-56 (AM~0624--260). Both galaxies have very close values of
O/H and M$_{\rm B}$. However, their M(HI) differ by a factor of 11.
Respectively, the ratios M(HI)/L$_{\rm B}$ differ for two LSBD galaxies by
a factor of $\sim$15.

In models of chemical evolution, LSB galaxies are usually treated as objects
with the lowest star formation rates (SFRs). This is related to very low
surface mass density and inefficient self-gravitation of discs in such
objects. Because of very low SFR, the related energy release is small, so that
newly produced heavy elements should retain in the parent galaxy. Thus, it
is expected that the chemical evolution of LSB galaxies should be well
described by the model of `closed-box', in which the ISM metallicity is
tightly related to the gas mass-fraction. Despite the amount of good data
on the ISM metallicity in LSB dwarfs to compare them with theoretical models
is still insufficient,
in some papers on LSB galaxies (e.g., de Naray et al. \cite{deNaray04}) no
significant deviations from the predictions of `closed-box' were found.
The example above gives an evidence that among the nearest LSB galaxies such
a tight relation is absent. Galaz et el. (\cite{Galaz02}) come to similar
conclusions from the NIR study of a large sample of LSB galaxies. This
implies that our understanding of chemical
evolution even for `the simplest'  galaxies requires more general models which
incorporate some additional significant factors.

\begin{table}[hbtp]
\caption{Main parameters of And~IV and ESO~489-56}
\label{tab:2_LSB}
\begin{tabular}{|l|l|l|c|} \hline  \hline
  Parameter                       & ~~And~IV   & ~~ESO~489-56 & Ref.    \\  \hline
B$_{\rm tot}$                     & ~~16.60    & ~~16.08      & 1,2     \\
A$_{B}$                           & ~~0.27     & ~~0.28       & 8     \\
V$_{\rm hel}$(\kms)               & ~~234      & ~~491        & 3,4     \\
Distance (Mpc)                    & ~~6.1      & ~~5.0        & 1       \\
M$_{\rm B}$                       & ~~--12.6   & ~~--12.63    & 1,2     \\
12+$\log$(O/H)                    & ~~7.49     & ~~7.49       & 5,6     \\
M(HI) (10$^8$M\sunn)              & ~~1.8      & ~~0.16       & 7,4     \\
M(HI)/L$_{\rm B}$ (M\sunn/L\sunn) & ~~13       & ~~0.85       & 7,4     \\ \hline   \hline
\MC{4}{l}{1. Karachentsev et al. (\cite{CNG}); 2. Pustilnik et al. (2008,} \\
\MC{4}{l}{in preparation); 3. Braun et al. (\cite{Braun03}); 4. Pustilnik, Martin }\\
\MC{4}{l}{(\cite{NRT}); 5. this work; 6. Ronnback, Bergvall (\cite{RB95}); } \\
\MC{4}{l}{7. Chengalur et al. (\cite{Chengalur07}); 8. Schlegel et al. (\cite{Schlegel98}). }\\
\end{tabular}
\end{table}

\subsection{Conclusions}

\begin{itemize}
\item
With multimode instrument SCORPIO at the BTA the good S-to-N spectra were
obtained of two brightest HII regions in And~IV, a dwarf LSB galaxy situated
at the distance of 6.1 Mpc.
\item
The oxygen abundances in HII-regions No.~3 and 4, determined by the classical
T$_{\rm e}$ method are consistent each to other. Their parameters
12+$\log$(O/H) (in old scale) are equal to 7.49$\pm$0.06 and 7.55$\pm$0.23,
respectively. The most reliable semi-empirical and empirical methods for
O/H estimates give values close to those calculated by the direct T$_{\rm e}$
method. In the new scale the accepted metallicity of And~IV corresponds to
12+$\log$(O/H)=7.50.
This galaxy appears one more a rare Local Volume LSBD galaxy
with very low metallicity (Z$\lesssim$Z\sunn/14).
\item
The comparison of global parameters of And~IV and the other LSBD in the
Local Volume, ESO~489-56, which have very close metallicities and optical
luminosities, reveals very large difference on their HI mass. This fact
indicates on difficulties of simple models of the chemical evolution of
LSB galaxies.
\end{itemize}

\begin{acknowledgements}
S.A.~Pustilnik and A.L.~Tepliakova acknowledge the support of this work
through the RFBR grant No. 06-02-16617. The authors are grateful to
I.D.~Karachentsev who brought this interesting Local Volume galaxy to their
attention. They are also very grateful to the referee N.G.~Guseva for useful
suggestions which allowed to improve the first version of the paper.
This research made use
of the NASA/IPAC Extragalactic Database (NED), which is operated by the Jet
Propulsion Laboratory, California Institute of Technology, under the contract
with the NASA.
\end{acknowledgements}

\end{document}